\documentclass[preprint,12pt]{elsarticle}

\usepackage{graphicx}
\usepackage{amssymb}
\usepackage{amsmath}
\usepackage{changepage}
\usepackage{color}

\usepackage{lineno}

\journal{Physica A: Statistical Mechanics and its Applications}

\begin{document}

\begin{frontmatter}


\title{The Role of Multiplex Network Structure in Cooperation through Generalized Reciprocity}


\author{Viktor Stojkoski$^{1}$}
\author{Zoran Utkovski$^{2,3}$}
\author{Elisabeth Andr{\'e}$^{4}$}
\author{Ljupco Kocarev$^{1,5}$}

\date{\today}

\address{
$^{1}$Academy of Sciences and Arts of the Republic of North Macedonia, P.O. Box 428, 1000 Skopje, North Macedonia}
\address{
$^{2}$Fraunhofer Heinrich Hertz Institute, Einsteinufer 37, 10587, Berlin, Germany}%
\address{
$^{3}$Faculty of Computer Science, University Goce Delcev Stip, P.O. Box 10-A, 2000 Shtip 2000, North Macedonia}%
\address{
$^{4}$Augsburg University, Universit{\"a}tsstr. 6a, 86159 Augsburg, Germany}%
\address{
$^{5}$SS. Cyril and Methodius University, Faculty of Computer Science and Engineering,  P.O. Box 393, 1000 Skopje, Macedonia}%

\begin{abstract}
Recent studies suggest that the emergence of cooperative behavior can be explained by \textit{generalized reciprocity}, a behavioral mechanism based on the principle of ``help anyone if helped by someone''. In complex systems, the cooperative dynamics is largely determined by the network structure which dictates the interactions among neighboring individuals. These interactions often exhibit multidimensional features, either as relationships of different types or temporal dynamics, both of which may be modeled as a ``multiplex'' network. Against this background, here we advance the research on cooperation models inspired by generalized reciprocity by considering a multidimensional networked society. Our results reveal that a multiplex network structure may enhance the role of generalized reciprocity in promoting cooperation, whereby some of the network dimensions act as a latent support for the others. As a result, generalized reciprocity forces the cooperative contributions of the individuals to concentrate in the dimension which is most favorable for the existence of cooperation.
\end{abstract}


\end{frontmatter}


\section{Introduction}

Ever since the pioneering work of Axelrod~\cite{Axelrod-1984} on the stability of direct reciprocity (tit-for-tat strategy) in a lattice structured iterated prisoner's dilemma, a lot of effort has been put into discovering how different reciprocal mechanisms for emergence of cooperation fare under various topological circumstances. In particular, in~\cite{Nowak-1993,Ohtsuki-2006,Santos-2006} the concept of network reciprocity was examined, and in~\cite{Peleteiro-2014} the conditions that lead to promotion of cooperation through indirect reciprocity in complex networks were explored. Other notable studies include the role of emotions~\cite{Martinez-2015,Moniz-2017} or punishment of defectors~\cite{Axelrod-1986,Mahmoud-2016,Perc-2017}, dynamical link formation through indirect reciprocity~\cite{Fu-2008}, and even considering geographical relations~\cite{li2017evolution}.

Under all these mechanisms cooperation evolves as an inherent feature of the competitiveness between the interacting individuals. Recent biological studies, however, suggest that cooperative behavior can also emerge and be sustained if it is based on \textit{generalized reciprocity}, a rule based on the principle ``help anyone if helped by someone''~\cite{Taborsky-2016}. In ~\cite{Rutte-2007,Bartlett-2006,Isen-1987} it was shown that cooperation may emerge under this mechanism as a consequence of the changes in the physiological state of the individuals caused by their positive experience from previous interactions. 

The first steps towards the development of a framework to study the role of a state-based generalized reciprocity update rule in networked societies were made in~\cite{Utkovski-2017}. In this work, the authors developed a simple model for pairwise interactions where individuals send cooperation requests to randomly chosen neighbors. The acceptance of the requests is stochastically determined by a sole variable called \textit{internal cooperative state} which reflects the individuals' current welfare. A distinctive characteristic of the model is that, in steady state, the simple decision rule promotes cooperation while, at the same time, prevents the individuals being exploited by their respective network environment.

While this and similar models shed valuable insights on the role that network topology plays in promoting cooperation, most of them have so far addressed only interactions on networks that are of one ``dimension'', ignoring possible multidimensional phenomena, i.e. multiplex network structures. This is an obvious drawback since real-life networks often exhibit heterogeneous properties within the edge structure that are of fundamental value to the phenomena present in the system~\cite{Kivela-2014}. For instance, in social network analyses the patterning and interweaving of different types of relationships are needed to describe and characterize social structures~\cite{Boorman-1976,White-1976}. In telecommunication networks, where control of the level of cooperation displayed by the nodes is needed to achieve efficiency~\cite{gajduk2014energy,chu2010opportunistic}, the physical edges are often ``sliced'' into multiple parts in order to support the requirement of different devices~\cite{Sherwood-2010,Nikaein-2015}. Even genetic and protein relations between organisms constructed in multiple ways are crucial for the analysis of their cooperative bindings~\cite{pandit2013genome,Stark-2006,DeDomenico-2015}. Another example is cooperation ecological systems where species interact in various ways~\cite{pilosof2017multilayer}.

To this end, here we extend the model introduced in~\cite{Utkovski-2017} to account for a multiplex network structure, with the aim to characterize the network cooperation dynamics under the assumption of a state-based behavioral mechanism rooted in generalized reciprocity\footnote{In this sense, we do not account for ``competition'' between strategies~\cite{Axelrod-1981}, nor assume evolutionary updates or imitation~\cite{Pacheco-2006}. Instead, we suppose that a form of a generalized reciprocity mechanism has evolved in the consciousness of the individuals, whose further dynamics is solely determined by the multiplex network structure.}.  
In our model the dimensions act as platforms which facilitate transactions between active members. The activity of the individuals is modeled by constraining their presence to one dimension per round, and by making them able to answer only to requests from that same dimension. This assumption is consistent with the random walk models on multiplex networks~\cite{de2014navigability}, and is justified in systems where the round duration is very short and/or when individuals have limited interaction capacities. The resulting mechanism, while preventing exploitation by other individuals, exhibits additional features that act as promoters of cooperation in a multiplex network structure. Specifically, by allowing for heterogeneous benefits and costs (i.e. different parameter values across different dimensions), we show that cooperation can survive in the observed dimension even if the cost exceeds the benefit, as long as there is another dimension which acts as a support (having benefit-to-cost ratio larger than one). This essential characteristic of the new model comes in contrast to one-dimensional networks where the benefit being larger than the cost is a prerequisite for cooperation. In particular, in a one dimensional network a benefit to cost ratio less than one implies that the cooperative individual has to carry a larger cost than the benefit the other individual receives, therefore it may be said that cooperation reduces the overall social welfare. In a multiplex network, we argue that this decrease in social welfare in a observed dimension is compensated by a large enough benefit to cost ratio in another dimension. Moreover, by introducing simple dynamics for the probability that an individual is present in a certain dimension, we show that, under a behavioral model based on generalized reciprocity, the cooperative contributions effectively concentrate to the dimension where most of their cooperative neighbors are also present. Based on these observations, we discuss connections to reinforcement learning, in particular to the model of Roth and Erev~\cite{roth1995learning} and extensions therein~\cite{camerer1999experience}.

The rest of the paper is organized as follows. In Section~\ref{sec:Background} we revisit the concept of generalized reciprocity and discuss the specifics of our state-based behavioral model. In Section~\ref{sec:model} we introduce the stochastic network interaction model, together with its deterministic counterpart. We also introduce more details about the state-based behavioral update. The exposition in Section~\ref{sec:Results} (Results) is organized in 3 subsections. In Section~\ref{sec:analytical_properties} we derive the conditions for the emergence and stability of cooperation with homogeneous parameters across the multiplex network dimensions. In Section~\ref{sec:heter-pars} we continue by relaxing this assumption and numerically examine the model properties with heterogeneous, i.e. nonidentical parameters across the network dimensions. This is done under the assumption of a random, but predetermined selection of the interaction dimension (i.e. fixed, predetermined probability of presence in a certain network dimension). In Section~\ref{sec:dimension_update} we analyse the cooperation dynamics under a modified rule for the dimension presence, according to which the individuals are free to adapt the probability of presence in each of the network dimensions as a function of their payoff (state-dependent probability of presence). The analysis is performed  across different types of random multiplex networks, with the aim to investigate the role of the network topology. In Section~\ref{sec:real-world-example} we test the model on an empirical dataset that describes the relationships between households of 75 Indian Villages (multidimensional networks)~\cite{banerjee2013diffusion}. The numerical simulations on this real-life example support the general conclusion that the multiplex network structure, combined with an adaptive dimension update rule, enhances network cooperation in the scenario with behavioral update based on generalized reciprocity. Section~\ref{sec:conclusion} (Conclusions) summarizes our findings and discusses possible directions for future work.

\section{Background}
\label{sec:Background}

Generalized reciprocity's roots lie within the concept of indirect reciprocity, a rule described as ``help someone who is helpful''~\cite{Boyd-1989,Nowak-2005-evol,Taborsky-2016}. In the literature it can also be found under the terms of ``upstream indirect reciprocity''~\cite{Boyd-1989} and ``upstream tit-for-tat''~\cite{Nowak-2007}. In biological systems indirect and generalized reciprocity are essentially two different ideas since the former requires advanced cognitive capabilities of the involved entities. In particular, with anonymous interactions, applying indirect reciprocity requires that each individual keeps tracks of the reputation of all potential co-interacting partners, whereas generalized reciprocity requires each individual to know only his own recent history. Due to the complexity, indirect reciprocity has only been documented in humans~\cite{Wedekind-2000}. On the other hand, real-life behavior based on generalized reciprocity has not only been found to be present in humans~\cite{Bartlett-2006,Baker-2014}, but has also been observed in many other organisms, including rats~\cite{Rutte-2007}, monkeys~\cite{Leimgruber-2014} and dogs~\cite{Gfrerer-2017}.   

The theoretical models that have been developed for the purpose of explaining the innate mechanism behind generalized reciprocity may roughly be divided into three main groups, though with possible overlaps. The first group encompasses deterministic behavioral update rules where the individuals base their decisions of whether to cooperate or not solely on the outcome of their last interaction~\cite{Pfeiffer-2005,vanDoorn-2012}. The models within the second group address the scenario where generalized reciprocity emerges as a result of a random walk in which the altruistic act of one individual initiates a chain (sequence) of similar acts across the network. Nevertheless, as stated in~\cite{Nowak-2007,Chiang-2011}, this mechanism by itself is not sufficient for the promotion of cooperation unless other mechanisms such as direct or network reciprocity are already in place.
The models within the third group address a behavioral update rule according to which the individual levels of cooperation are adjusted on the basis of an internal state reflecting the individuals' general well-being (i.e. fitness). This adjustment takes place over time as a result of interactions with other individuals~\cite{Barta-2011,Utkovski-2017}. 

The specifics of our model are such that it may be categorized as being in the intersection between the first and the third group. In particular, it shares the state-based behavioral update with the models from the third group. On the other hand, by approximating the stochastic interactions by a deterministic model, as done in~\cite{Pfeiffer-2005}, our model is related with the models in the first group. Nevertheless, differently from the other deterministic approaches where the strategy choice was assumed to be binary (either fully cooperate or defect), our model can be placed in a continuous iterated Prisoner dilemma framework~\cite{Ranjbar-2014b,Ranjbar-2014a}, with the note that the choice of the particular strategy (from the continuous set of possible strategies) in each round is determined by the individual's state (reflecting its accumulated payoff, i.e. well-being). 

Compared to the classical (i.e. one-dimensional) random graph model, the multiplex network structure offers additional degrees of freedom which render the extension of the one-dimensional models nontrivial. While the evolution of cooperation in multiplex networks has been addressed in the context of other types of reciprocity~\cite{Gomez-2012,Battiston-2017,deng2018multi}, our interaction model and the state-based behavioral update yield interesting implications on the cooperative behavior in these networks. Similar to~\cite{Utkovski-2017}, the accent here is on the role of the network structure and the thereby related neighborhood importance index, here generalized to account for the different temporal interaction model due to the multiplex structure. 

\section{Model Description}
\label{sec:model}

\subsection{Network interaction model}
\label{sec:mplexcoop_Model}

We consider a population of $N$ individuals whose relations are modeled as a connected multiplex network, defined as the triplet $\mathcal{G} \left( \mathcal{N}, \mathcal{E}, \mathcal{L} \right)$, where $\mathcal{N}$ (the set of nodes) corresponds to the set of individuals, $\mathcal{E} \subseteq \mathcal{N} \times \mathcal{N}$ is the set of edges that describes the relationships between pairs of individuals, and $\mathcal{L}$ is the set of $L$ properties that can be attributed to the edges and which define the dimensions of the network. Formally, a dimension can be defined as the graph $\mathcal{G}^{\left[l\right]} \left( \mathcal{N}, \mathcal{E}^{\left[l\right]} \right)$ in which $\mathcal{E}^{\left[l\right]}$ is the subset of $\mathcal{E}$ having the property $ l \in \mathcal{L}$. Each dimension is given via an $N \times N$ adjacency matrix $\mathbf{A}^{\left[l\right]}$, where the $ij$-th entry $A^{\left[l\right]}_{ij} \in \{0,1\}$ between pairs of individuals $i, j \in \mathcal{N}$ ($ A^{\left[l\right]}_{ij}=1$ indicating neighborhood relation, i.e. $\left(i, j \right) \in \mathcal{E}^{\left[l\right]}$). 

The interactions between the individuals are modeled as follows: in each round $t$, each individual $i$: 
\begin{enumerate}
 \item randomly chooses a dimension $l$ where it will be present in that round;
 \item sends a cooperation request to a randomly (on uniform) chosen individual $j$ from its neighborhood in the $l$-th dimension, $j\in\mathcal{N}^{\left[l\right]}_i$;
\item upon selection, if individual $j$ is present in the the $l$-th dimension in round $t$, it receives the request and cooperates with probability $\mathrm{p}_j(t)$ representing the individual's internal cooperative state at round $t$; When cooperating, individual $j$ pays a cost $c^{\left[l\right]} > 0$ for individual $i$ to receive a benefit $b^{\left[l\right]} > 0$.
\end{enumerate}
Given this interaction model, the random payoff of individual $i$ at round $t$ may be characterized as
\begin{align}
\mathrm{y}_i(t)&=\sum_l v^{\left[l\right]}_{i}(t) \left[ b^{\left[l\right]} v^{\left[l\right]}_{j}(t) \mathrm{x}_{j}(t) -c^{\left[l\right]} \mathrm{x}_i(t)\sum_{k\in\mathcal{N}^{\left[l\right]}_i} \rho^{\left[l\right]}_{k}(t) v^{\left[l\right]}_{k}(t) \right].
\label{eq:mplexcoop_payoff}
\end{align}
In~(\ref{eq:mplexcoop_payoff}), $v^{\left[l\right]}_{i}(t)$ is the $i$-th outcome of an $L$ dimensional categorical variable parametrized by $B^{\left[l\right]}_{i}(t)$, which itself is a random variable describing the probability that $i$ is present in layer $l$ in round $t$. The selected index from the neighborhood of $i$ is a random variable uniformly distributed on the set $\mathcal{N}^{\left[l\right]}_i$, $j\sim \mathrm{U}(\mathcal{N}^{\left[l\right]}_i)$; $\mathrm{x}_j(t)$, $j=1,\ldots, N$, are Bernoulli random variables, each with parameter $\mathrm{p}_j(t)$; $\rho^{\left[l\right]}_k$ is a Bernoulli random variable with parameter $1/d^{\left[l\right]}_k$, where $d^{\left[l\right]}_k$ is the degree of individual $k$ in dimension $l$, $d^{\left[l\right]}_k=\sum_h A^{\left[l\right]}_{kh}$; the term $\sum_{k\in\mathcal{N}^{\left[l\right]}_i} \rho^{\left[l\right]}_{k}(t)$ captures the random number of individuals (neighbors of $i$ in $l$) which send a cooperative request to $i$ in dimension $l$ during round $t$. 

\subsection{Deterministic approximation} 
\label{sec:mplexcoop_Deterministic_Model}
We approximate the stochastic model~(\ref{eq:mplexcoop_payoff}) by a deterministic model in which the random variables are substituted with their respective expectations
\begin{align}
\mathrm{y}_i(t)&=\sum_l B^{\left[l\right]}_{i}(t) \left[ b^{\left[l\right]} \sum_j \frac{A^{\left[l\right]}_{ij}}{d^{\left[l\right]}_i} B^{\left[l\right]}_{j}(t) \mathrm{p}_{j}(t) - c^{\left[l\right]} z^{\left[l\right]}_i(t)  \mathrm{p}_{i}(t) \right].
 \label{eq:deterministic_mplexcoop_compact}
\end{align}
The term $z^{\left[l\right]}_i(t)$ in (\ref{eq:deterministic_mplexcoop_compact}) is defined as
\begin{align}
z^{\left[l\right]}_i(t)=\sum_j \frac{A^{\left[l\right]}_{ji}}{d^{\left[l\right]}_j} B^{\left[l\right]}_{j}(t),
\end{align} 
is the temporal extension of the neighborhood importance index discussed in~\cite{Utkovski-2017}, in dimension $l$. This quantity acts as a local centrality measure of an individual, with individual $i$ being more ``important'' in the addressed dimension if it has many neighbors which are at the same time also present in that dimension, and the neighbors themselves have few neighbors. In our model of interactions, this individual would be called upon rather frequently in the studied dimension.   

\begin{figure*}[t]\centering
 \begin{adjustwidth}{0in}{0in}
\includegraphics[width=13cm]{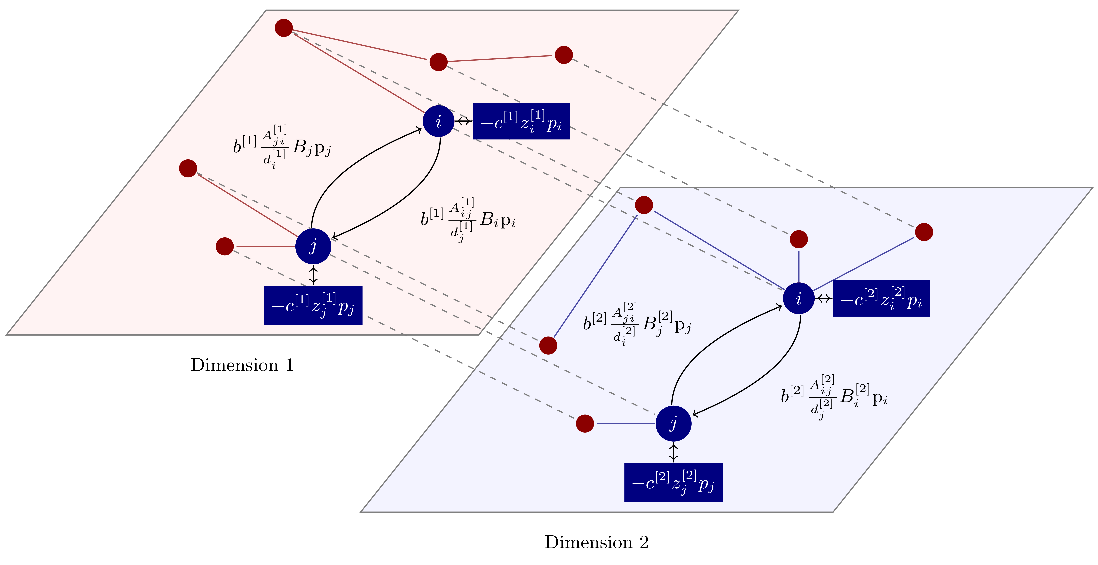}
\end{adjustwidth}
\caption{An example for the deterministic interactions on a two dimensional network between two individuals $i$ and $j$ (colored in dark blue). For illustrative purposes we exclude the round notation. Filled colored edges indicate neighborhood relation in the corresponding dimension, whereas dashed lines are links to the same individual in the other dimension.}
\label{fig:mplex_coop_eg}
\end{figure*}

The motivation to use the deterministic model~(\ref{eq:deterministic_mplexcoop_compact}) is that it captures the long-term behavior of the stochastic model, i.e. provides reliable approximation of its steady state behavior. Numerical simulations of the stochastic model in the long run suggest that the stochastic variables can indeed be approximated by their respective expectations, without affecting the long-term network behavior, thus justifying the approximation. In Fig.~\ref{fig:mplex_coop_eg} we display an example for the deterministic interactions between two individuals that are placed on a two-dimensional network.

We also define the aggregate time-dependent neighborhood importance index $Z_i(t)$ of individual $i$ as
\begin{align}
Z_i(t) &= \frac{\sum_l B^{\left[l\right]}_{i}(t) z^{\left[l\right]}_i(t)}{\sum_l B^{\left[l\right]}_{i}(t) \sum_j \frac{A^{\left[l\right]}_{ij}}{d^{\left[l\right]}_i}B^{\left[l\right]}_{j}(t)}.
\end{align}
As we will see in more detail in Section~\ref{sec:Results} (Results), the probability distribution of the quantity $Z(t)$ across the individuals crucially determines the global cooperative behavior in the network. This quantity, which is a form of aggregated centrality measure in the multiplex network setting, critically reflects the role of the network topology on the cooperation dynamics in our interaction model. We elaborate on this in more detail in Section~\ref{sec:Results} where we address multiplex networks generated from Erdos-Renyi (ER) and Barabasi-Albert (BA) random graph models, as well as a real-world example of a social network describing relationships between households in Indian villages~\cite{banerjee2013diffusion}.  

With the above, we write~(\ref{eq:deterministic_mplexcoop_compact}) in a more compact (vector) form as  
\begin{align*}
\mathbf{y}(t) = \mathbf{\Theta}(t)\cdot\mathbf{p}(t), 
\end{align*}
where $\Theta_{ii}(t)= -\sum_l c^{\left[l\right]} B^{\left[l\right]}_{i}(t) z^{\left[l\right]}_i(t)$, and $\Theta_{ij} (t) = \sum^{\left[l\right]} b^{\left[l\right]} B^{\left[l\right]}_{i}(t) \frac{A^{\left[l\right]}_{ij}}{d^{\left[l\right]}_i} B^{\left[l\right]}_{j}(t)$, for $i\neq j$.

\subsection{Behavioral update rule}

We study a synchronous update rule, based on the accumulated payoff of the in $i$ by round $t$, $\mathrm{Y}_{i}(t)=\mathrm{Y}_{i}(t-1)+\mathrm{y}_{i}(t)$, with $\mathrm{Y}_{i}(0)$ being the initial condition and $\mathrm{y}_{i}(0)=0$.  
The cooperative state of $i$ at round $t+1$ is defined as 
\begin{align}
\mathrm{p}_i(t+1)=\mathrm{f} \left[\mathrm{Y}_{i}(t)\right],
\label{eq:update}
\end{align} 
where we assume that the function $\mathrm{f}:\mathbb{R}\rightarrow [0,\:\:1]$ is increasing. A plausible choice which reflects real-world behavior is the logistic function 
\begin{align*}
\mathrm{f}(\omega)=\left[1+e^{-k(\omega-\omega_0)}\right]^{-1},
\end{align*}
where the parameters $k$ and $\omega_0$ define the steepness and the midpoint of the function. 

The justification behind using one state variable for each individual lies in the fact argued in Section~\ref{sec:Background}, i.e. that generalized reciprocity has been predominantly found among individuals with lower cognitive capabilities. Nevertheless, the ability to choose the dimension presence may indirectly provide the needed flexibility for the individuals to exhibit different cooperation intensities in different dimensions. 

Moreover, we point out that it is straightforward to extend the model to account for an asynchronous behavioral update, where in each step $t$ individual $i$ updates its probability of cooperation with probability $u$. In that case, however, the steady state cooperative behavior of each individual does not depend on the choice of $u$, unlike the case for network reciprocity~\cite{allen2017asynchronous}.


\section{Results}
\label{sec:Results}

\subsection{Analytical properties of the model} 
\label{sec:analytical_properties}
Here we characterize the main properties of the model in steady state. Hereby, we distinguish between two types of results: 1) results with homogeneous parameters, defined as identical benefit and costs across network dimensions, $b^{\left[l\right]} = b$ and $c^{\left[l\right]} = c$ for all $l \in \mathcal{L}$; and 2) results with heterogeneous parameters, i.e. the more general case when we allow for different values for the benefits and costs across network dimensions. We remark that the proofs for the properties follow directly from applying the framework presented in references~\cite{stojkoski2018cooperation,Utkovski-2017}, nevertheless for concreteness in the exposition we present them. 

The following property holds in general. 

\textbf{1.~Robustness~to~exploitation.} In steady state, the individuals may be attributed to two disjoint sets, $\mathcal{W} = \left\{ w \in \mathcal{N} : \mathrm{y}^*_w=0  \right\}$ and $\mathcal{S} = \left\{ s \in \mathcal{N} : \mathrm{y}^*_s>0  \right\}$, based on the steady state payoff $\mathrm{y}^*_i$. The individuals in $\mathcal{S}$, which we refer to as ``strong individuals'', are characterized by $\mathrm{p}^*_i=1$, while the individuals in $\mathcal{W}$, called ``weak'' may take both values $\mathrm{p}^*_i=1$ and $\mathrm{p}^*_i<1$, depending on the network parameters. Hence, there are two sets of relations that have to be satisfied 
\begin{align}
0& =\sum_l B^{\left[l\right]*}_{i} \left[ b^{\left[l\right]} \sum_j \frac{A^{\left[l\right]}_{ij}}{d^{\left[l\right]}_i} B^{\left[l\right]}*_{j} \mathrm{p}_{j}^* - c_l z^{\left[l\right]*}_i  \mathrm{p}_{i}^* \right], \:\: i\in \mathcal{W}\nonumber\\
\mathrm{y}^*_i &= \sum_l B^{\left[l\right]*}_{i} \left[ b^{\left[l\right]*} \sum_j \frac{A^{\left[l\right]}_{ij}}{d^{\left[l\right]}_i} B^{\left[l\right]*}_{j} \mathrm{p}_{j}^* - c^{\left[l\right]} z^{\left[l\right]*}_i  \mathrm{p}_{i}^* \right], \:\: i\in \mathcal{S}.
\label{eq:optimization_model_simplified}
\end{align}  
Note that the sets $\mathcal{W}, \mathcal{S}$, the steady state values $\mathrm{p}_i^*,\:i\in\mathcal{W}$ and $B^{\left[l\right]*}_{j},\:i\in\mathcal{N},\:l\in\mathcal{L}$ and the constants $\mathrm{y}^*_i, \:i\in \mathcal{S}$ are unknown.

\textit{Proof:} The update rule (\ref{eq:update}) yields the following set of iterative equations for $i=1,\ldots,N$
\begin{align*}
\mathrm{p}_i(t+1) &= \mathrm{f}\left(\mathbf{\mathrm{Y}}_i(t-1)+\mathbf{\Theta}_i(t)\cdot\mathbf{p}(t)\right),
\end{align*}
where $\mathbf{\Theta}_i(t)$ is the $i-$th row of $\mathbf{\Theta}(t)$. 
In equilibrium it has to be fulfilled 
\begin{align*}
\mathrm{p}^*_i&=\mathrm{f}\left(\mathrm{f}^{-1}\left(\mathrm{p}^*_i\right)+\mathbf{\Theta}^*_i\mathbf{p}^*\right),
\end{align*}
for $i=1,\ldots,N$. By applying the inverse map we get 
\begin{align}
\mathrm{f}^{-1}\left(\mathrm{p}_i^*\right)&=\mathrm{f}^{-1}\left(\mathrm{p}_i^*\right)+\mathbf{\Theta}^*_i\mathbf{p}^*.
\label{eq:steady_state_individual_inverse}
\end{align}
The above requires $\mathrm{y}^*_i\doteq\mathbf{\Theta}^*_i\mathbf{p}^*=0$ which further implies $\mathrm{p}^*_i = 0$, unless either $\mathrm{p}_i^*=1$ (i.e. $\mathbf{\mathrm{Y}}_i^*=\mathrm{f}^{-1}\left(\mathrm{p}_i^*\right)=\infty$), or $\mathrm{p}_i^*=0$ (i.e.$\mathbf{\mathrm{Y}}_i^*=-\infty$). 

It is easy to verify that if there exists $i$ such that $\mathrm{p}^*_i=0$, then the same is true for all $i\in\mathcal{N}$. Indeed, when $p^*_i=0$, then from (\ref{eq:deterministic_mplexcoop_compact}) and since $\sum_l B^{\left[l\right]*}_{i} = 1$, it must hold that either: 1) $\mathrm{y}^*_i>0$, or: 2) $\mathrm{p}^*_j=0$ for all $j$ in the neighborhood of $i$, $j\in \bigcup_{l\in \mathcal{L} } \mathcal{N}^{\left[l\right]}_i$. The condition 1 implies $\mathrm{p}^*_i=1$, which is a contradiction. The condition 2 yields $\mathrm{p}^*_i=0$ for all $i\in\mathcal{N}$ by repeating the same argument to the nodes in the neighborhood of $i$, until all individuals are reached. We note that this case is also covered by the requirement $\mathbf{\Theta}^*_i\mathbf{p}^*=\mathbf{0}$, with the solution $\mathbf{p}^*=\mathbf{0}$. 
Hence, an equilibrium fulfills $\mathbf{p}^*\in \mathbf{0}\cup (0\:\:1]^N$ and is thereby characterized by non-negative steady state payoffs $\mathrm{y}^*_i\geq 0$. \hfill $\blacksquare$

The following properties hold only for a multiplex network with homogeneous parameters. We relax this assumption in the numerical analysis performed in the following sections.

\textbf{2.~Existence~of~cooperation.} A necessary condition for existence of cooperators (individuals with $\mathrm{p}^*_i>0$) is $b/c \geq 1$.

\textit{Proof:} Note that the total network payoff can be written as
\begin{align} \label{eq:network-payoff}
\sum_i \mathrm{y}^*_i&= \sum_l \left(b^{\left[l\right]} - c^{\left[l\right]} \right) \sum_i B^{\left[l\right]*}_{i} z^{\left[l\right]*}_i \mathrm{p}^*_{i}. 
\end{align}
It is easy to show that, $b/c < 1$ implies $\mathrm{p}^*_i = 0$ for all $i\in\mathcal{N}$. Indeed, if there exists $i$ such that $\mathrm{p}^*_i>0$, then the total steady state network payoff is strictly negative, implying that there is some $i$ for which $\mathrm{y}^*_i < 0$ (contradiction). Hence, the necessity of $b/c\geq 1$ for existence of cooperation.\hfill $\blacksquare$

\textbf{3.~Promotion~of~cooperation.} When $b/c>1$, we observe the steady state probabilities are strictly greater than $0$, $\mathrm{p}_i^*>0$ for all $i\in \mathcal{N}$.

\textit{Proof:} By contradiction. If there exists $i$ such that $p^*_i=0$ then, as already discussed, it must hold that $\mathrm{p}^*_i=0$, for all $i\in\mathcal{N}$. This, however, would yield a total network payoff $\sum_{i}\mathrm{y}^*_i=0$, which contradicts (\ref{eq:network-payoff}). \hfill $\blacksquare$

\textbf{4.~Sufficient~condition~for~existence~of~strong~individuals.} When $b/c>1$, there is always at least one strong individual in the network. 

\textit{Proof:}
This follows from the observation that when $b/c>1$ the right-hand-side of (\ref{eq:network-payoff}) is strictly greater than zero, which implies that there is at least one $i$ for which $\mathrm{y}^*_i > 0$ and $\mathrm{p}^*_i = 1$. \hfill $\blacksquare$

\textbf{5.~Necessary~condition~for~the~existence~of~strong~individuals.} A necessary condition for existence of strong individuals, (individuals with $\mathrm{p}^*_i = 1$), is $Z^*_i \leq b/c$.

\textit{Proof:}
The proof follows directly by substituting $\mathrm{p}^*_i = 1$ and the fact that $b^{\left[l\right]} \sum_j \frac{A^{\left[l\right]}_{ij}}{d^{\left[l\right]}_i} B^{\left[l\right]*}_{j} \geq b^{\left[l\right]} \sum_j \frac{A^{\left[l\right]}_{ij}}{d^{\left[l\right]}_i} B^{\left[l\right]*}_{j} \mathrm{p}^*_j$. \hfill $\blacksquare$

\textbf{6.~Full~network~cooperation.} The condition $b/c \geq Z^*_{max}$, where $Z^*_{max}$ is the largest neighborhood importance index in the graph, $Z^*_{max} = \max_i Z^*_i$, is both necessary and sufficient for all individuals to be strong. 

\textit{Proof:} We note that the proof that $\mathrm{p}_i^*=1$, $\forall i\in \mathcal{N}$, implies $b/c \geq Z^*_{max}$, follows directly from \textit{property 3}. To prove the converse, we use contradiction. We first define $\mathrm{p}_{min}=\inf \mathrm{p}_i^*, \: i\in \mathcal{N}$, and set $b/c$ to be greater than one (since $b/c>1$ is prerequisite for cooperative behavior). 
Now, let us assume that the converse is not true, that is $ b/c \geq Z^*_i$ for all $i$, and there exists some $i$ such that $\mathrm{p}^*_i < 1$. Under this assumption, for all $i \in \mathcal{W}$  we would have:

\begin{align*}
\mathrm{y}_i^*&= b\sum_l B^{\left[l\right]*}_{i} \sum_j \frac{A^{\left[l\right]}_{ij}}{d^{\left[l\right]}_i} B^{\left[l\right]*}_{j} \mathrm{p}_{j}^* - c \sum_l B^{\left[l\right]*}_{i} z^{\left[l\right]*}_i  \mathrm{p}_{i}^* \\
& \geq b\sum_l B^{\left[l\right]*}_{i} \sum_j \frac{A^{\left[l\right]}_{ij}}{d^{\left[l\right]}_i} B^{\left[l\right]*}_{j} \mathrm{p}_{j}^* -c \frac{b}{c}\sum_l B^{\left[l\right]*}_{i} \sum_j \frac{A^{\left[l\right]}_{ij}}{d^{\left[l\right]}_i} B^{\left[l\right]*}_{j}  \mathrm{p}^*_i.
\end{align*}
Since, if $p^*_i< 1$, $y^*_i = 0$, this implies
\begin{align}
 \mathrm{p}^*_{i} \geq  \frac{\sum_l B^{\left[l\right]*}_{i} \sum_j \frac{A^{\left[l\right]}_{ij}}{d^{\left[l\right]}_i} B^{\left[l\right]*}_{j} \mathrm{p}_{j}^*}{\sum_l B^{\left[l\right]*}_{i} \sum_j \frac{A^{\left[l\right]}_{ij}}{d^{\left[l\right]}_i} B^{\left[l\right]*}_{j}} \geq \mathrm{p}_{min} .
 \label{eq:pmin}
\end{align}
For all $i$ satisfying $b/c > Z^*_i$,  (\ref{eq:pmin}) holds with strict inequality, whereas those $i'$ for which $b/c = Z^*_{i'} $ must satisfy $\mathrm{p}^*_{i'} = \mathrm{p}_{min}$. This, however, can hold if and only if the individuals corresponding to these indices are only linked to each other in each dimension and have the same degree in them, i.e. form a connected component. In that case $Z^*_{max} = 1 = b/c$ which contradicts the assumption $b/c > 1$. Hence, the converse must also be true, which concludes the proof. \hfill $\blacksquare$

\subsection{The role of heterogeneous parameters}
\label{sec:heter-pars}

We continue the analysis by relaxing the assumption of homogeneous parameters, and consider the situation where each dimension $l$ has its own benefit $b^{\left[l\right]}$ and cost $c^{\left[l\right]}$. Since our goal is to examine the effect of heterogeneous parameters, we develop a null model in which the probability for dimension presence is uniform in every round.

For this case, we compare two different multiplex networks each composed of two dimensions (the results can be easily generalized to networks with more dimensions). In particular, the first multiplex network type represents a natural generalization of the Erdos-Renyi (ER) random graph, while the second is a multiplex version of the Barabasi-Albert (BA) preferential attachment graph. In an ER random graph an edge between two individuals has a fixed probability of being present, independently of the other edges. As a consequence the degree follows a Poisson distribution. On the other hand, a BA graph is constructed by a dynamical process in which in each step a new individual is introduced, and this individual makes connections to other individuals that are already in the graph with probability proportional to their degree, thus ending up with a power law degree distribution. More details about the algorithms used for generating these multiplex networks can be read in references~\cite{bianconi2013statistical,nicosia2013growing,nicosia2014nonlinear}.

In what follows, we will consider networks which consist of 100 individuals, and where the average degree in each dimension is 8. Since the correlations between the edges in different dimensions should play a prominent role in determining the steady state level of cooperation we are going to study three different scenarios. In the ER graph, we will examine the possibility of overlapping edges, i.e. situations i) when there is no edge overlap in different dimensions, ii) when half of the edges overlap, and iii) when all edges overlap. In the BA graph we study the cases where the degree correlation $\rho$ generated through preferential attachment is i) negatively correlated, ii) there is no correlation, and iii) is positively correlated between the dimensions. Formally we measure the correlation between the degrees in separate dimensions through the Pearson correlation coefficient, i.e
\begin{align}
\rho_{12} &= \frac{\sum_i \left( d^{\left[1\right]}_i - \langle d^{\left[1\right]}\rangle \right) \left( d^{\left[2\right]}_i - \langle d^{\left[2\right]}\rangle \right) }{  \sqrt{\sum_i \left( d^{\left[1\right]}_i - \langle d^{\left[1\right]}\rangle \right)^2} \sqrt{\sum_i \left( d^{\left[2\right]}_i - \langle d^{\left[2\right]}\rangle \right)^2}}.
\label{eq:pearson_correlation}
\end{align}
In equation (\ref{eq:pearson_correlation}) $\langle d^{\left[l\right]}\rangle$ denotes the average degree in dimension $l$.

\begin{figure*}[t!]
\includegraphics[width=12cm]{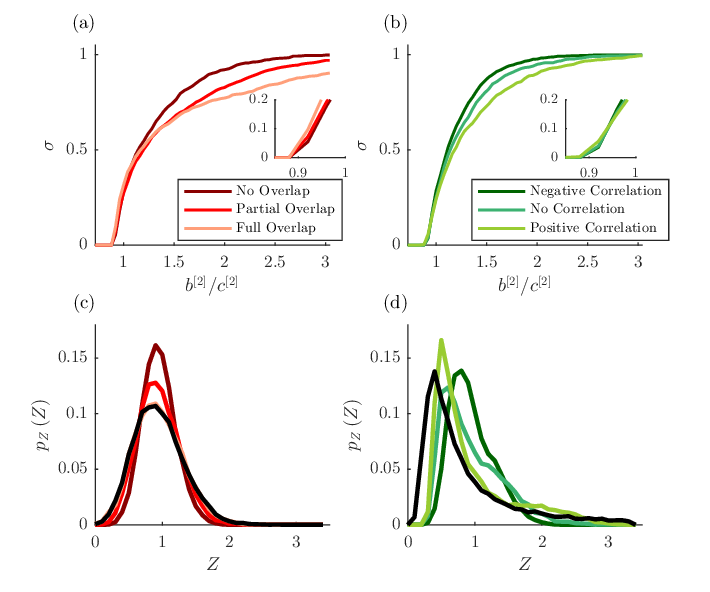}
\caption{\textbf{(a-b)} Fraction of strong individuals $\sigma$ as a function of the benefit to cost ratio in the second dimension, $b^{\left[2\right]}/c^{\left[2\right]}$, for a sample of the random graphs that we study, while $b^{\left[1\right]}/c^{\left[1\right]} = 1.08$. \textbf{(a)} ER multiplex networks with uniform update. \textbf{(b)} BA multiplex networks with uniform update. \textbf{(c-d)} Probability density function for the steady state $Z$ index for the same random graphs averaged across 1000 realizations and the corresponding one-dimensional graphs (in black). \textbf{(c)} ER multiplex networks. \textbf{(d)} BA multiplex networks. \textbf{(c-d)} The black curve illustrates the distribution of $Z$ in the one-dimensional representation of the corresponding random graph. All networks have 100 individuals and average degree 8. }
\label{fig:heterogen-pars}
\end{figure*}

The results are depicted in Fig.~\ref{fig:heterogen-pars}. Panels~(a)-(b) respectively show the evolution of the fraction of strong individuals for the multiplex network generated by the ER and BA multiplex networks while the benefit to cost ratio in the second dimension, $b^{\left[2\right]} / c^{\left[2\right]}$ is varied, whereas the benefit to cost ratio in the first dimension is kept constant ($b^{\left[1\right]} / c^{\left[1\right]} = 1.08$). 
In the case of ER networks, it can be noticed that full edge overlap is better for the level of cooperation displayed when $b^{\left[2\right]} / c^{\left[2\right]}$ is low (see the inset plot of Fig.~\ref{fig:heterogen-pars}~(a)), while no edge overlap promotes more cooperation for higher values of the benefit to cost ratio in the second dimension. Similarly, positive degree correlation in the BA networks leads to higher fraction of strong individuals for small benefit to cost ratios(inset plot of Fig.~\ref{fig:heterogen-pars}~(a)), while negative degree correlation is better for supporting cooperation when $b^{\left[2\right]} / c^{\left[2\right]}$ is high.

We point out that due to the uniform probability of dimension update there is a symmetric relationship between changes in the benefit to cost ratios in both dimensions and the fraction of strong individuals, i.e. a change in $b^{\left[l\right]} / c^{\left[l\right]}$ in one dimension has the same effect as a change in the other dimension. Any slight adjustment in the probabilities only changes the symmetry towards one of the dimensions, which means that the results should not be gradually altered. Hence, we can use the properties derived in the previous section as a starting point for the inspection of the results.

This reduces the analysis to studying the steady state distribution of the index $Z$. We note that, according to our interaction model, higher values of the index imply more frequent cooperation requests, and hence, lower incentives for cooperation. For this purpose in panels~(c)-(d) of Fig.~\ref{fig:heterogen-pars} we plot the typical probability density function (PDF) for the $Z$ index for the same random multiplex networks where the dimension presence is given by a uniform probability by averaging the index across 1000 network realizations. There, in black, we also depict the PDF of $Z$ for the one dimensional ER and BA networks\footnote{Note that the full overlap multiplex ER graph coincides with the one-dimensional representation.}. For the ER graphs, it can be seen that the exclusion of overlapping edges effectively increases the mode of the distribution. Since the average level of cooperation at the point where cooperation begins to exist is determined by the left tail of the distribution, the networks with lower modes, and thus larger left tails, should be able to promote more cooperation. Contrastingly, the fatness of the right tail leads to higher thresholds for displaying full network cooperation. In a similar fashion, we notice that by decreasing the correlation between the degrees in the multiplex BA graph, the right tail of $Z$ decreases, and therefore the lower threshold for full cooperation.

Finally, in the inset plots of  Fig.~\ref{fig:heterogen-pars} (a)-(b) we observe that the inclusion of a second dimension leads to significantly lower threshold for existence of cooperation in the system (which in the one dimensional case is $b / c > 1$~\cite{Utkovski-2017,stojkoski2018cooperation}). This implies that the other dimension acts as a support for existence of cooperation even if the original dimension does not allow it. This is a result of the fact that the negative payoffs from the original dimension are compensated with positive payoffs from the supporting dimension. If at least one individual receives higher steady state payoff from the supporting dimension than the loss in the original, then cooperation will persist. This is an important implication to the emergence of cooperation in systems where all dimensions of the network can not be observed and the environment is not suited for cooperation due to not increasing the social welfare, while the phenomena is still detected.

\subsection{The role of dynamics in the dimension update rule}
\label{sec:dimension_update}

Predetermined presence is a plausible assumption for systems where the flow between dimensions is constrained and individuals are not allowed to develop beliefs about which dimensions generate higher payoffs to them. A more realistic case would be to allow for dynamics in the probability that individual $i$ is present in dimension $l$ in round $t$. While this can be modeled by introducing Markov transition rates for moving from one dimension to another, or even adding memory rates to the movement based on the experience in the previous rounds, here we consider a simpler update rooted in the same generalized reciprocity rule that was used for the internal state update. 

Concretely, we consider an update based on the accumulated payoff in the dimension, 
\begin{equation}
\mathrm{Y}^{\left[l\right]}_{i}(t)=\mathrm{Y}^{\left[l\right]}_{i}(t-1)+\mathrm{y}^{\left[l\right]}_{i}(t),
\label{eq:payoff_propensity}
\end{equation}
with $\mathrm{Y}^{\left[l\right]}_i(0)$ being the initial condition and $\mathrm{y}^{\left[l\right]}_{i}(0)=0$. In our model the updated probability of presence of individual $i$ in dimension $l$ is given by the softmax function
\begin{align}
B^{\left[l\right]}_{i}(t+1) = \frac{\exp\left( \mathrm{Y}^{\left[l\right]}_{i}(t) \right) }{ \sum_{m} \exp\left( \mathrm{Y}^{\left[m\right]}_{i}(t) \right)}.
\label{eq:dimension_update}
\end{align}
We remark that the described rule is similar in spirit to the famous Roth-Erev reinforcement learning algorithm for strategies in extensive form games~\cite{roth1995learning}. In fact, based on (\ref{eq:payoff_propensity}) and (\ref{eq:dimension_update}), an analogy with more general reinforcement learning models can be established~\cite{camerer1999experience,jost2014reinforcement}. The connection is provided by interpreting the act of presence of individual $i$ in dimension $l$ of the multiplex network as a selection of a strategy $S_l$ (from a set of $L$ preselected strategies), where the strategy selection is applied with probability $B^{\left[l\right]}_{i}$. In this context, the total payoff $\mathrm{Y}^{\left[l\right]}_{i}$ in the network dimension $l$ is analogous to the propensity to play strategy $S_l$. We note that the here applied  rule~(\ref{eq:dimension_update}) is different from the one introduced in the original Roth-Erev learning model, according to which strategy $l$ is selected with probability $B^{\left[l\right]}_{i}(t+1) = \frac{\mathrm{Y}^{\left[l\right]}_{i}(t)}{ \sum_{m} \mathrm{Y}^{\left[m\right]}_{i}(t)}$. Specifically, it can be considered as a special case of a more general reinforcement learning scheme in which the
probabilities for players to choose certain actions are taken
from a general Gibbs-Boltzmann distribution
\begin{equation}
B^{\left[l\right]}_{i}(t+1) = \frac{\exp\left( \lambda \cdot \mathrm{Y}^{\left[l\right]}_{i}(t) \right) }{ \sum_{m} \exp\left( \lambda \cdot \mathrm{Y}^{\left[m\right]}_{i}(t) \right)}.
\label{eq:Gibbs_Boltzmann}
\end{equation}
In~(\ref{eq:Gibbs_Boltzmann}) $\lambda$ plays the role of ``inverse temperature'' in statistical physics, and captures the trade-off between exploitation ($\lambda=\infty$), i.e. greedy
learning in which only the action with the highest propensity is taken, and exploration ($\lambda=0$), meaning that all actions are equally probable. In many reinforcement learning problems, the key is to find a value of $\lambda$ that achieves a reasonable trade-off between exploitation and exploration in the model of question. In our model, the selection of this parameter would critically determine the transient and evolutionary behavior in the system when, for example, some dimensions are erased. We expect that in that case selecting the parameter $\lambda$ towards the ``exploration mode'' would provide certain robustness to such events. The exact quantification of these effects under this scenario, together with the role of the network topology, is out of the scope of this manuscript. However, it represents an interesting direction for future work.

The advantage of the suggested update is that it can be very easily implemented since the individual only needs to know the probability $B^{\left[l\right]}_{i}(t)$ in the current round for all dimensions and the received payoff $\mathrm{y}^{\left[l\right]}_{i}(t)$ from them.

We point out that the steady state solution of equation~(\ref{eq:dimension_update}) is not defined if more than one $\mathrm{Y}^{\left[l\right]}_{i}$ tends to infinity. This never happens as long as each individual experiences different dynamics when the dimensions are considered as separate networks. Another thing worth emphasizing is that the resulting system has $\left( L - 1 \right)N$ degrees of freedom, and, hence, complex behavior is unavoidable. Therefore, in the comparative statics as starting points for the probability that an individual is present in a certain dimension we consider real numbers whose values are comparable to the steady state of the preceding benefit to cost ratios. In the beginning, at the lowest benefit to cost ratios, the starting point is set to be equal among all individuals and dimensions. 

\begin{figure*}[t!]
\includegraphics[width=12cm]{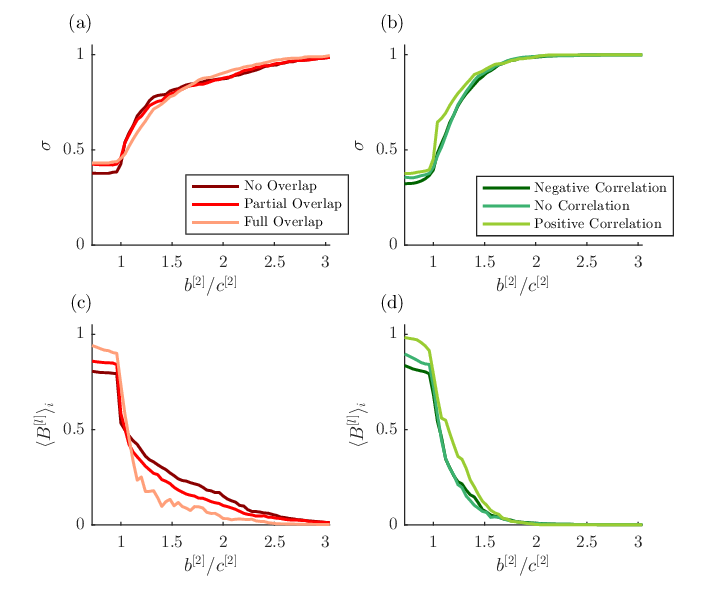}
\caption{\textbf{(a-b)} Fraction of strong individuals $\sigma$ as a function of the benefit to cost ratio in the second dimension, $b^{\left[2\right]}/c^{\left[2\right]}$, for a sample of the random graphs that we study, while $b^{\left[1\right]}/c^{\left[1\right]} = 1.08$. \textbf{(a)} ER multiplex network with generalized reciprocity update. \textbf{(b)} BA multiplex network with generalized reciprocity update. \textbf{(c-d)} For the same graphs, the fraction of individuals $\langle B^{\left[l\right]*} \rangle$ which in the steady state are present in the first dimension.}
\label{fig:dimension-update}
\end{figure*}

The results for the same networks as in the previous section are shown in Fig.~\ref{fig:dimension-update}. Panels~(a)-(b) depict the fraction of strong individuals as a function of the benefit to cost ratios. On the one hand, we observe that the global level of cooperation displayed for low $b^{\left[2\right]} / c^{\left[2\right]}$ ratios is increased by a large amount when compared to the predetermined probability for dimension presence. On the other hand, we notice that for larger benefit to cost ratios, the overall level of displayed cooperation is not consistent in terms of performance. More precisely, there are situations in which it is decreased (e.g. the no-overlap ER network), and there are situations in which it is increased (e.g. the positive correlation BA network) when compared to predetermined presence.

This aggregate behavior can be explained by looking at (c)-(d) of Fig.~\ref{fig:dimension-update}, where we display the fraction of individuals which in steady state are present in the first dimension as a function of the same parameters. Obviously, the dimension in which the individuals are always present in steady state is not always the same, i.e. it is dispersed among the individuals depending on the network topology and parameters. This is a key feature of the model since it implies that the dimension update rule forces the individuals to accommodate their presence towards the dimension where they either carry the smallest burden to cooperate or where most of their cooperative neighbors are present. As such, when coupled with the generalized reciprocity state update rule (\ref{eq:update}), the dimension update rule facilitates the promotion of cooperation in the system (in the sense that it promotes the existence of individuals with $\mathrm{p}^*_i > 0$). 

It is worth mentioning that the fact that individuals may accommodate their presence to the dimension where most of their cooperative neighbors are present indicates that full unconditional cooperation is not guaranteed to be achieved in an easier fashion. In particular, the dimension update rule may force an individual in steady state to be present in a dimension which is less suited for its personal gain than some predetermined rule just because most of the other individuals are present in that dimension. Exactly this may be the cause for the lower level of cooperation exhibited in the no-overlap ER network. Due to the manner in which the edges are constructed in this graph, it may happen that some individuals to have high neighborhood importance index in the first dimension but low in the other. For these individuals, a uniform predetermined rule would imply that they would be able to generate a higher payoff by interacting in the dimension where they are less burdened. However, none of their neighbors with lower value of $Z$ would be better of if they are present in it due to lower benefit to cost ratio and, more importantly, due to being less forced to answer to a cooperation request. As a consequence, these group of individuals accommodate their presence in the second dimension, and the individuals with high neighborhood importance index in it are also forced to be present in it. 

\subsection{Real-world examples}
\label{sec:real-world-example}

As a means to provide an intuitive example for the experimental application of the model we utilize social network data that describes relationships between households in Indian Villages~\cite{banerjee2013diffusion}. In this dataset there are a total of 75 villages (networks) each represented through 12 separate dimensions. Since, there is a significant overlap in the way the separate dimensions are constructed (see~\cite{banerjee2013diffusion} for a detailed description), here we consider only 4 dimensions that describe essentially disparate types of social interactions. In the first dimension, the links represent relationships between households who help each other with making decision, the second and third, respectively, describe the borrowing interactions of money and kerosene and rice. Finally, the fourth dimension are the medical advice relations.

For estimation purposes, we exclude households that have no neighbors in at least one of the studied dimensions. Thus we end up studying 75 different multiplex networks each consisting of 4 dimensions and on average $126.45$ individuals (with a standard deviation of $43.59$). More detailed summary statistics are given in Table~\ref{table:village-summary}.

\begin{table}[h]
\centering
\caption{Villages networks summary statistics}
\label{table:village-summary}
 \begin{adjustwidth}{-2cm}{0cm}
\begin{tabular}{|l|c|c|c|c|}
\hline
\textbf{Dimension} & \textbf{Avg. degree} & \textbf{Med. degree} & \textbf{Avg. clustering} & \textbf{Avg. path} \\
\hline \hline
Help with Decision  & 3.21 (0.48) & 2.70 (0.56) & 0.16 (0.06) & 4.35 (1.28) \\ \hline
Borrow money & 3.76 (0.73) & 3.35 (0.78) & 0.21 (0.06) & 4.14 (1.22) \\ \hline
Borrow rice and kerosene & 4.26 (0.71) & 3.69 (0.69) & 0.19 (0.07) & 3.76 (0.84) \\ \hline
Give medical advice & 3.82 (0.68) & 3.48 (0.72) & 0.26 (0.05) & 4.46 (1.32) \\ \hline
\end{tabular}
\flushleft{Note: Standard deviations in brackets.}
\end{adjustwidth}
\end{table}

\begin{figure}[t!]
\includegraphics[width=12cm]{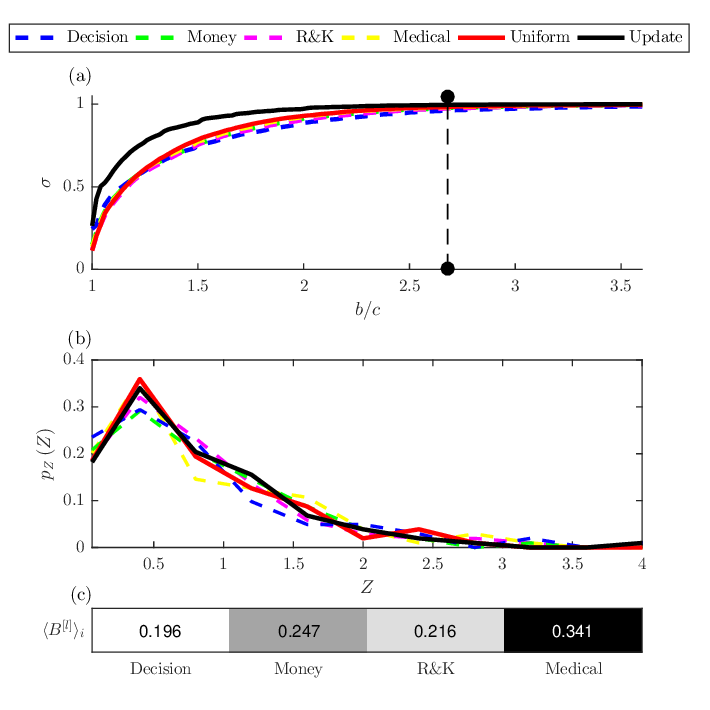}
\caption{\textbf{(a)} Average fraction of unconditional cooperators $\sigma$ as a function of $b/c$ for the multiplex networks with dimension update rule (black), with uniform update (red), and for the individual networks: help with a decision (blue), borrow money (green), borrow rice and kerosene (magenta) and give medical advice (yellow). The dashed vertical line indicates the threshold for full network cooperation in the update rule case.
\textbf{(b)} Average distribution of the index $Z$ for the networks. The illustrated distribution for the networks in which the dimension update rule is at force is estimated with the steady state probabilities averaged across estimations.
\textbf{(c)} Average fraction of steady state presence $\langle B^* \rangle_i$ when there is a generalized reciprocity update rule for it.}
\label{fig:villages-res}
\end{figure}

For easier interpretation of the results, in the numerical estimations we set homogeneous benefits and costs. In Fig.~\ref{fig:villages-res} panel (a) we plot the average fraction of strong individuals as a function of the benefit and cost ratio $b/c$ across all multiplex networks with and without the dimension update rule. There, we also depict the same variable when each dimension is considered as a separate network. In general, we observe that the dynamics of the examples in which the dimension update rule is at force supports most cooperation, followed by the uniform update rule. In this particular example, the individual dimensions behave as worst promoters of cooperation based on generalized reciprocity. 

Similar to the previous results, these are also best explained if we look at the steady state distribution of the index $Z$ (panel~(b) of Fig.~\ref{fig:villages-res}). In particular, it can be seen that the average distribution for $Z$ in the multiplex network with dimension update rule has tails that are less fat than the other example structures. Clearly, this is a result of the individuals being able to accommodate to the dimension where they receive highest payoff, differently from the other structures that are compared. As evidenced in panel~(c) of Fig.~\ref{fig:villages-res}, where we show the average steady state probability for dimension presence (averaged across individuals and estimations), in steady state the individual presence distribution for the dimensions is well diversified. While most of the individuals choose to be present in the Medical advice dimension, there are some individuals that favor the other dimensions. Consequently, the mass of the distribution of $Z$ is driven towards one. In other words, Property 5. is easily satisfied for most individuals solely by assuming $b/c > 1$ and Property 6. is reached with lower $b/c$ ratio. On the other hand, all other distributions exhibit fatter tails, and hence require higher benefit to cost ratio in order for full cooperation to appear. This is especially true for the help with a decision dimension when it is considered as a separate network. Evidently, in it there is a small group of individuals that receive cooperation requests way more often than they send. The fat tail of $Z$ in the help with a decision dimension can be a direct result of the fact that this dimension has on average the lowest clustering coefficient. Concretely, many of the possible triads of edges are not formed, which in turn leads to excessive burden to a particular group of individuals for which the edges in the possible triad exist.

As a final remark, we state that the individuals in the investigated system are groups of humans and as such are probably able to develop higher cognitive senses for competition and cooperation than simply following a generalized reciprocity rule. Nevertheless, this does not reduce the significance of the results since they serve as a demonstration that the inclusion of multiple dimensions together with the update rule can guide a real-life system to a higher level of cooperation.

\section{Conclusions}
\label{sec:conclusion}

The emergence of cooperation in complex networks precludes the existence of a specific behavioral mechanism and a particular network interaction structure~\cite{Nowak-2006five}. This interaction structure often exhibits multidimensional features such as relationships of different types or temporal dynamics.

Against this background, we studied the cooperation dynamics under a behavioral mechanism based on generalized reciprocity, in a network consisting of multiple dimensions, each modeled by a random graph. The model, which is a generalization of the one introduced in~\cite{Utkovski-2017}, provides new insights on the role of the network structure on the promotion of cooperation in complex multidimensional networks. In particular, we demonstrated that a multidimensional structure may support cooperation within the individual network dimensions, even when the benefit-to-cost ratio in the considered dimensions is below the threshold required for cooperation (when observed in isolation). This observation may explain the existence of cooperation in systems where cooperative behavior is observed even though it does not increase social welfare -- a latent structure (i.e. other dimensions) may exist that acts as a support to the observed dynamics.

We also discussed the connection between the studied behavioral mechanism in the multidimensional network and reinforcement learning, by interpreting the act of presence of individuals in the dimensions of the multiplex network as a selection of a strategy (from a predefined set of strategies), where the strategy selection is applied in relation to individuals' internal state. In this context, we introduced a simple and intuitive rule for modeling the individuals' interactions in the different dimensions (i.e. their presence across dimensions). The experiments were performed both on an multidimensional extension of the random graph models, and on a real-life dataset. As a general observation, the cooperative contributions of the individual individuals concentrate in the dimension which is most favorable for the existence of cooperation. 

An interesting direction for future work is the study of more general behavioral mechanisms in the spirit of the ``exploration vs. exploitation'' discussion in reinforcement learning. In this context, it will be valuable to study not only the steady-state (in the sense of evolutionary behavior), but also the transient dynamics in the network. Another fruitful topic is generalizing the multiplex model to other truly cooperative games that cannot be sustained in absence of other behavioral mechanism. As a final note, the model can also be used as a starting point in the examination of network formation based on generalized reciprocity, where the neighborhood of each individual in each dimension can be seen as the possible final outcome of a rewiring process that is determined by the dimension update rule. 

\section*{Acknowledgement}

This research was supported in part by DFG through grant ``Random search processes, L\'evy flights, and random walks on complex networks''.


\end{document}